\documentclass[10pt,journal,letterpaper]{IEEEtran}

\usepackage{graphicx,cite,epsfig,amssymb,amsmath,amsfonts,multicol,subfigure,mathtools,bm,mathrsfs,setspace,xcolor}
\usepackage{multirow}
\usepackage{cite}

\begin{document}
\title{\huge A Quasi-Doppler Method for Doubling Transmission Efficiency Through Two Orthogonal Directions}
\author{ 
	Bingli Jiao, \IEEEmembership{Senior Member,~IEEE}

\thanks{Bingli Jiao  is with the Department of Electronics and Peking University-Princeton University Joint Laboratory of Advanced Communications Research, Peking University, Beijing 100871, China (email: jiaobl@pku.edu.cn).}
}

\maketitle

\begin{abstract}

Inspired by the anisotropy of Doppler effect with wave propagations, we propose a new method to leverage one information symbol serving two users located in two geometrically orthogonal directions.  Specifically in broadband wireless communication, we use multiple antennas with the proposed signal switching method to emulate a moving emission source and yield the frequency shift, referred to as Quasi Doppler effect, which is converted to the discrete phase modulation.  Further, using this discrete modulated phase can adjust the phase of one transmit symbol in achieving two different phases in different directions.  The modulation mechanism is explained through theoretical derivations with the analysis on the performance robustness in the application-scenarios of crossroads having small geometric deviations. In contrast to the use of conventional symbols, this approach can double the transmission efficiency which is confirmed by our simulations results.

\end{abstract}

\begin{IEEEkeywords}
Quasi-Doppler, spatially perpendicular channel, phase-modulated signal.
\end{IEEEkeywords}

\IEEEpeerreviewmaketitle

\section{Introduction}
\IEEEPARstart{A}s has been known, in conventional wireless communications, sending one information symbol from the transmitter to the receiver constitutes a transmission process over air interface.  Apart from the desired receiver, the transmitted symbol can be interference to the other receivers, except for broadcasting services. 

This work proposes a new method that enables one information symbol to serve, over air interface, two users located in geometrically orthogonal directions as shown in Figure~1.  This scenario provides our research motivation to double transmission efficiency in the orthogonal geometry with the parallel- and perpendicular channel. The physical mechanism, allowing one symbol to appear with two different phases in the two directions, can be essentially attributed to the use of the spatial anisotropy of Doppler effect, which has been categorized into several sorts in general \cite{Jackson1998}[2][3].   

Our research starts from considering the Doppler effect on its emitted electromagnetic waves in the plane wave model and the observations from different angles with respect to the moving direction of the emission source.  

When a harmonic signal is emitted, the Doppler effect can be expressed mathematically by 
\begin{eqnarray}
	\begin{array}{l}\label{1}
	\Delta f(\theta) = f^{'}-f=\frac{v_x}{\lambda}\cos\theta,   \ \ \
	\end{array}
\end{eqnarray}
where $\Delta f$ is the Doppler frequency shift, $f^{'}$ is the frequency observed, $f$ is the frequency of the emission, $v_x$ is the speed of the source, $\theta$ is the angle between directions of the observation and the source's moving, $\lambda$ is the wavelength. 

In a two-dimensional Cartesian coordinate model,  we use $\Delta f_x$ and $\Delta f_y$ to express $\Delta f(0)$ and $\Delta f(\pi/2)$ for showing that $\Delta{f}_y = 0 $ holds no matter what value $\Delta{f}_x $ is. This indicates that any effects resulted from $\Delta f_x$ will be nullified in the direction of y-axis. 

Now, we work on $\Delta f_x$ with Maxwell equation\cite{Jackson1998} by
\begin{eqnarray}
\begin{array}{l}\label{2}
R_x(t)= Ae^{j2\pi f t-j(kx_0-kv_xt)}, \\ 
\end{array}
\end{eqnarray}
where $j=\sqrt{-1}$, $R_x(t)$ is the received wave of a receiver located at x-axis for $x>0$, $A$ is the received amplitude, $x_0$ is the initial position of the receiver, and $k=2\pi/\lambda$ is a constant.   

\begin{figure}[!t]
	\centering
	\includegraphics[width=0.35\textwidth]{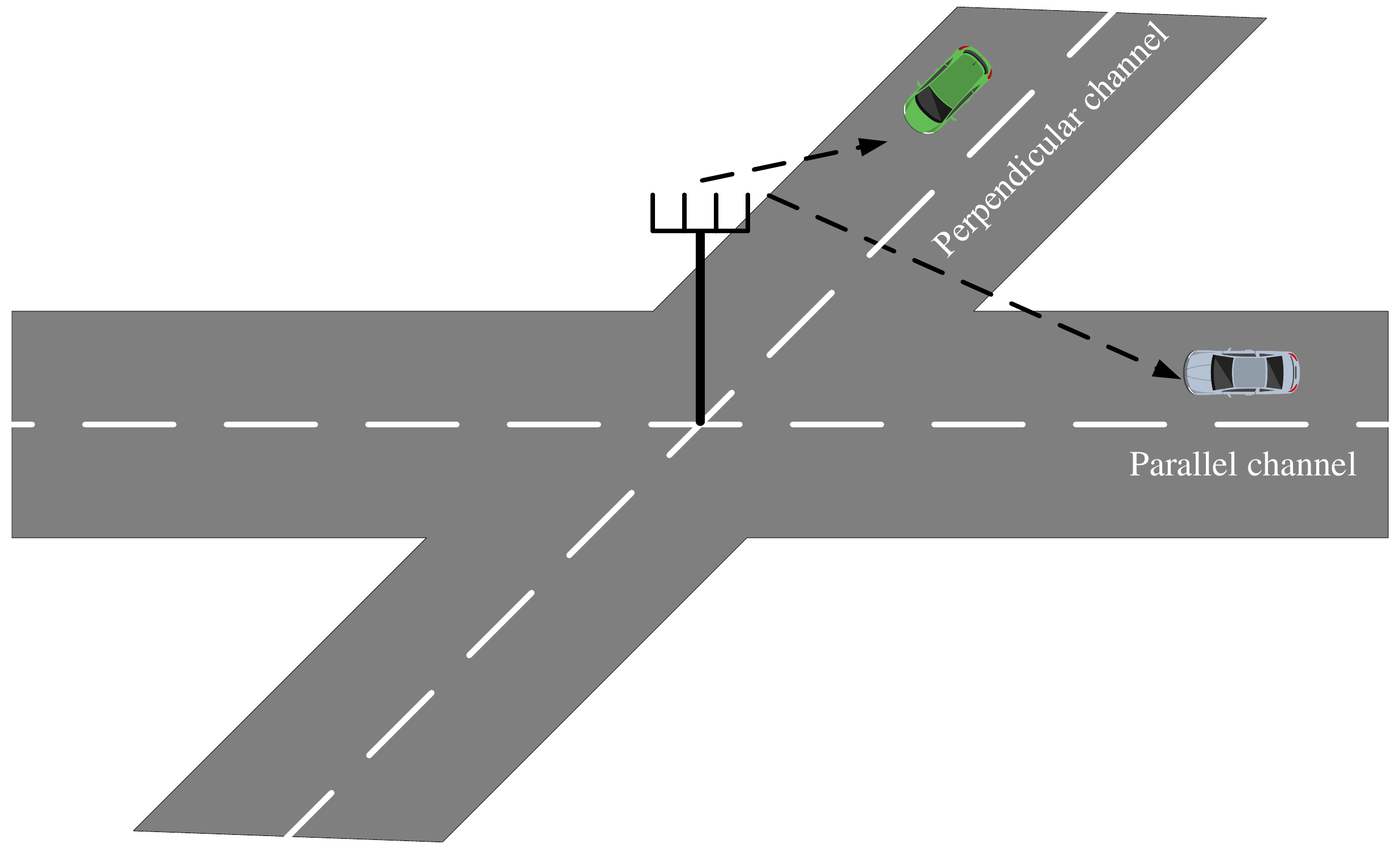}
	\caption{ Multiple antennas at the crossroads working through  the parallel- and perpendicular channel.}
	\label{Application Scenario}
\end{figure}

The Doppler frequency shift is found at $v_xk/2\pi = v_x/\lambda$ again in \eqref{2}.  However, because $v_xt$ is actually the position the emission source arrives at time $t$, the frequency shift can be regarded as a continuous phase modulation with respect to the position of the emission source.  This can be more clearly seen at the baseband level by removing the carrier frequency $f$, i.e.,
\begin{eqnarray}
\begin{array}{l}\label{3}
\hat{R}_x(x)=Ae^{jkx(t)-j\phi_0},  \ \ \ 
\end{array}
\end{eqnarray}
where  $\hat{R}_x(x)$ is the received baseband signal, $x(t)=v_x t$ denotes the position of the emission source at time $t$, $\phi_0= kx_0$ is the initial phase.  

It is noted that $kx(t)$ is a modulated phase that does not appear in direction of y-axis.  
     
\section{Signal Modulation}
In this section, we propose a method in which the transmitter consists of a signal source and multiple antennas to transmit information symbols to the two receivers $\hat{R}_x$ and $\hat{R}_y$ located in two geometrically orthogonal directions, where the parallel and perpendicular channels are defined respectively, as shown in Fig.2. 

\begin{figure}[!t]
	\centering
	\includegraphics[width=0.35\textwidth]{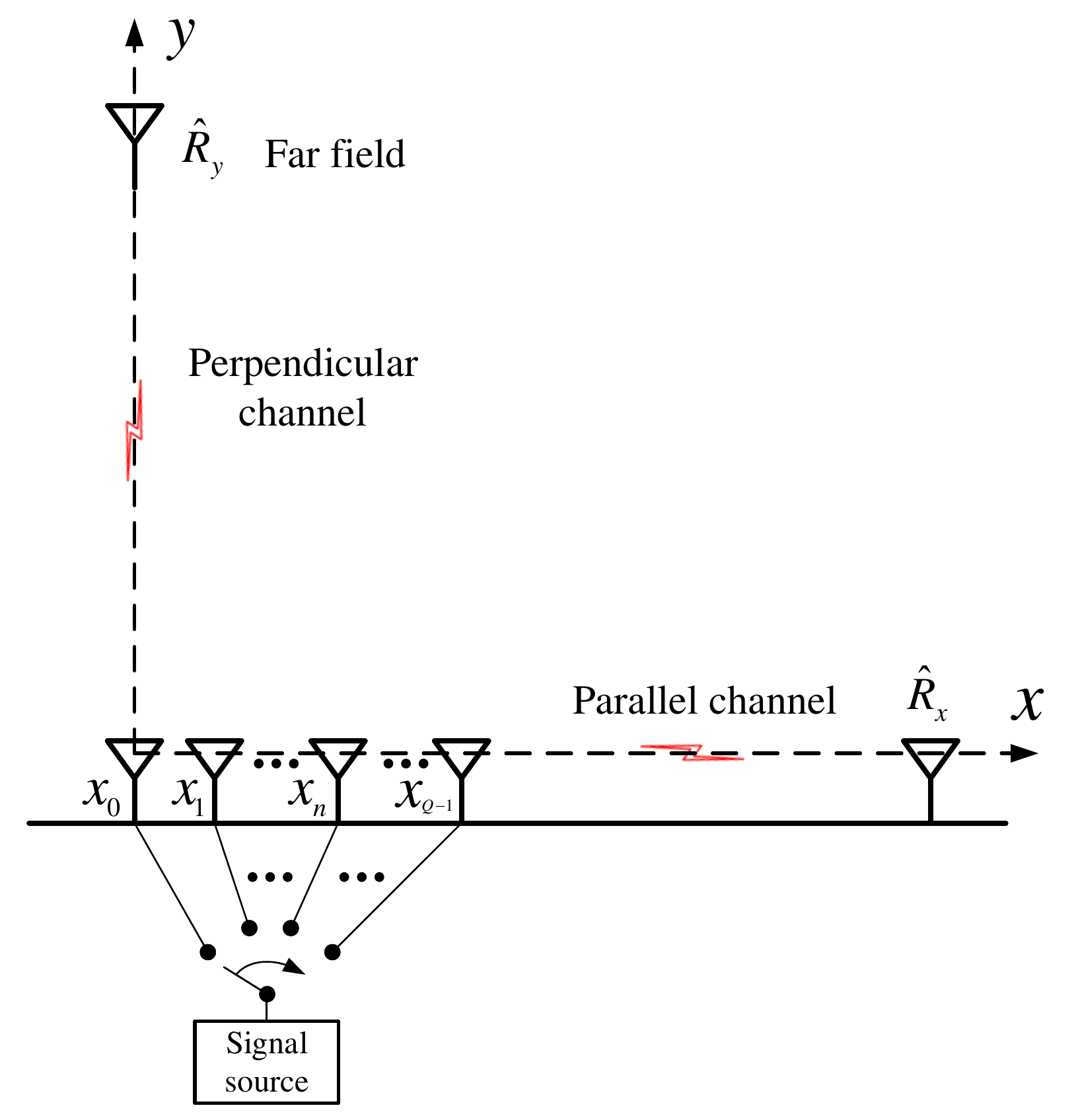}
	\caption{The transmitter architecture of the QD method and the ideal geometric model.}
	\label{PhD}
\end{figure}

\subsection{Quasi-Doppler Effect}
Let us use a harmonic signal, in the plane wave model, to explain the mechanism of generating Doppler effect by using multiple antennas to replace the moving of the emission source.  These antennas are uniformly distributed along x-axis at $x_0 = 0,~x_1 = d,\cdots,x_q=qd,\cdots$, till $x_{Q-1} = (Q-1)d$ as shown in  Fig.2, where $d$ is the distance between two adjacent antennas and $Qd= \lambda$ is required for limiting the geometric range of the multi antennas.  

To emulate the harmonic signal transmission of the Doppler source moving at speed $v_x$, we switch the signal source sequentially onto each antenna $x_0,~x_1,\cdots,x_q,\cdots, x_{Q-1}$, where the time of connecting the signal source and each antenna is set to $T=d/v_x$ and the switching time, i.e., the time of hopping between any two antennas, is assumed to be zero. The received signal through the parallel channel can be written as
\begin{eqnarray}
\begin{array}{l}\label{4}
\hat{R}_x(t)= h_xe^{jkv_x T \lfloor t/T\rfloor+j\phi_0} + n_0, \ \ \
\end{array}
\end{eqnarray}
where $\hat{R}_x(t)$ is the received baseband signal, $v_x = d/T$ is the speed of emulated moving, $\lfloor \cdot \rfloor$ is the floor operator, $h_x$ is a complex channel response, $\phi_0$ is an initial phase and $n_0$ is the Gaussian noise.

This equation shows the Doppler frequency shift at $kv_x/2\pi$, which is referred to as the Quasi Doppler (QD) effect, because the time variable behaves in a step manner with $\lfloor t/T \rfloor$.  

Similar with the transformation from \eqref{2} to \eqref{3}, assuming $\phi_0=0$, we can transfer the QD frequency shift into desiccate phase modulations, according to the positions of those antennas, as expressed by
\begin{eqnarray}
\begin{array}{l}\label{5}
\hat{R}_x(x_q)=h_xe^{jk x_q+j\phi_0} + n_0, 
\end{array}
\end{eqnarray}
where $x_q$ represents the position of $q^{th}$ antenna, $q = 0,1,\cdots,Q-1$.  

In the following discussions, ``Antenna q'' stands for the antenna's position at position $x=x_q$.  It is found that the QD effect yields the discrete phase modulation instead of the continuous one in \eqref{3}.  

Further, switching the signal source from antennas $q'$ to antenna $q$ can provide a phase modulation at $\Delta {\phi} = k(x_q-x_{q'}) = 2\pi( x_q-x_{q'})/\lambda$.   Thus, by assuming $\phi_0=0$, the transmission at antenna $q$ can be regarded as a QD phase of 
\begin{eqnarray}
\begin{array}{l}\label{6}
\phi_q =2\pi x_q  /\lambda,  
\end{array}
\end{eqnarray}
where $\phi_q$ is defined as the QD phase, and $x_q$ is the position of the transmit antenna.  

For preparing a signal constellation for the transmission, $x_q$ should be pre-designed by
\begin{eqnarray}
\begin{array}{l}\label{6-05}
x_q = \phi_q \lambda/2\pi    
\end{array}
\end{eqnarray}
for $q=1,2,....,Q-1$, according to the required $\phi_q$ in the signal constellation.  

In the communication scheme, the assumption of $\phi_0 = 0$ is reasonable, because the channel characterization permits.  

\subsection{Joint Symbol Modulation}
The transmission issue arises with the use of one information symbol to provide two desired phases that appear through the parallel channel and the perpendicular channel.  
To be specific, $\phi_x$ and $\phi_y$ denote these two desired phases. The joint signal modulation is explained in the following two steps.  

At the first step, the QD transmitter uses the conventional modulation at its signal source to produce an information symbol as  
\begin{eqnarray}
\begin{array}{l}\label{7}
S_y = \sqrt{E_s}e^{j\phi_y},  
\end{array}
\end{eqnarray}
where $S_y$ , $E_s$ and $\phi_y$ are the modulated symbol, the symbol energy and the desired phase for the perpendicular channel, respectively. 

Secondly, for providing the desired phase, i.e., $\phi_x$, through the parallel channel, the produced symbol in \eqref{7} should be switched onto an appropriate antenna to complete the symbol transmission.  The QD phase adds its contribution by        
\begin{eqnarray}
\begin{array}{l}\label{8}
S_x = \sqrt{E_s}e^{j(\phi_y+ \phi_q)}, 
\end{array}
\end{eqnarray}
and letting $\phi_y+ \phi_q = \phi_x$ leads to 
\begin{eqnarray}\label{9}    
\phi_q = \begin{cases}
\begin{split}
&\phi_x - \phi_y,  \quad &\phi_x \ge \phi_y,\\
&\phi_x +2\pi - \phi_y, \quad &\phi_x <\phi_y,
\end{split}
\end{cases}
\end{eqnarray}
where $\phi_q$ is the QD phase.  Further, by using \eqref{6-05}, we find the transmit antenna $q$ for $x=x_q$.
  
Because the QD phase. i.e., $\phi_q$, does not show up through the perpendicular channel, the symbol transmission can be summarized in form of 
\begin{eqnarray}
\begin{array}{l}\label{10}
\hat{R}_\beta = h_\beta S_\beta + n_0 =  h_\beta\sqrt{E_s}e^{j\phi_\beta} + n_0, 
\end{array}
\end{eqnarray}
where $\hat{R}_\beta$ and $h_\beta$ are the received signal and the channel gain factor, respectively, with $\beta=x$ and $y$ indicating the signal received through the parallel- and perpendicular channel, respectively.   
  
Based on the above derivations, we conclude that the QD method provides a possibility to double the transmission efficiency in the orthogonal geometry model, because one jointly modulated symbol can satisfy the needs of two receivers.   

\section{On the Performance}
The purpose of this section is to explore the application issues in the scenarios of crossroads, whereat the ideally orthogonal- and deviated geometric model are taken into the theoretical analysis and the simulations as well.  

\subsection {Theoretical Analysis}
\begin{figure}[!t]
	\centering
	\subfigure[]{
		\centering
		\includegraphics[width=0.38\textwidth]{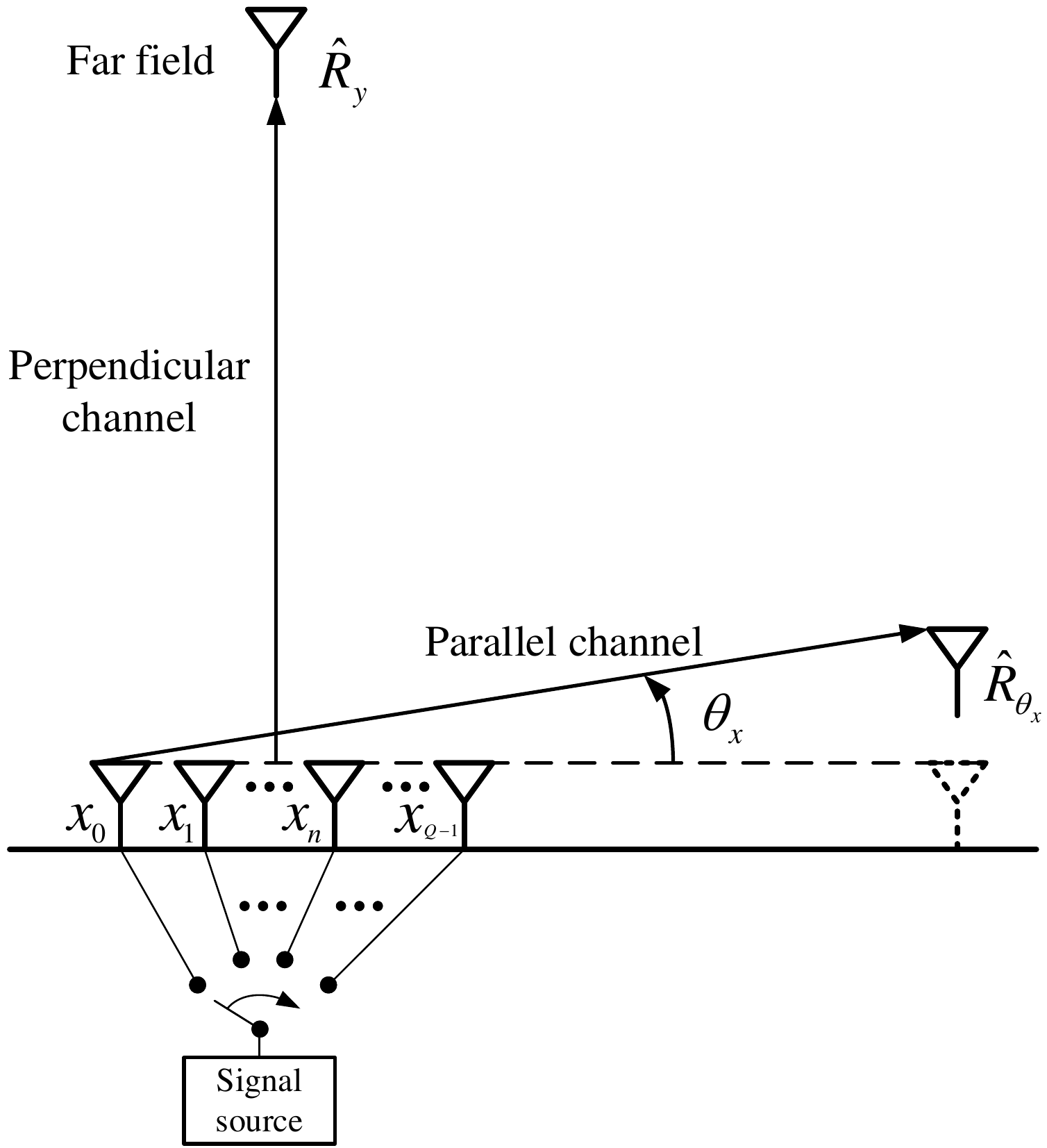}
		\label{fig3a}
	}
	\subfigure[]{
		\centering
		\includegraphics[width=0.38\textwidth]{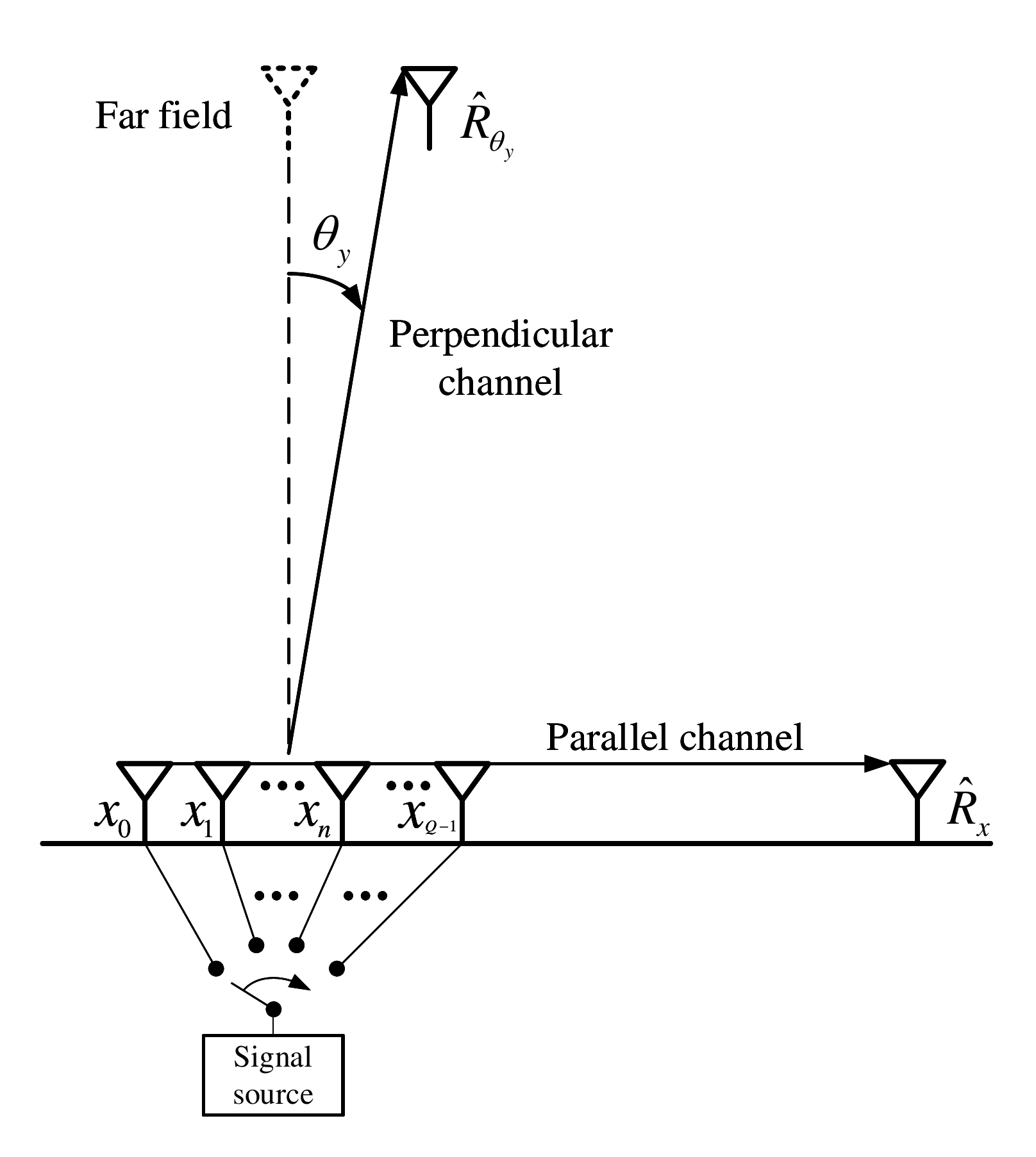}
		\label{fig3b}
	}
	\caption{Deviated geometric model:  an deviated angel with the parallel channel in (a) and that with perpendicular channel in (b). }
	\label{fig3}
\end{figure}

In addition to the ideal geometric model, we define the deviated geometric model for the scenarios, where the orthogonality is slightly deviated, as shown in Fig. \ref{fig3}, and work on the cases of the parallel- and the perpendicular channel separately for showing the discrepancies. 

As can be expected, in comparison with the ideal geometry, the deviated angles can cause some phase offsets as analysed below. 

Assuming that there is a small angle deviation $\theta_x$ between x-axis and the line of sight from the transmit antenna to the receiver $\hat{R}_x$ as shown in Fig. 3(a),  we can find that the received $\phi_y$ remains unchanged in \eqref{8} and, however, the QD phase can change to (see the Appendix) 
\begin{equation}\label{11}
{\phi}_{q}' = k x_q\cos\theta_x =  2\pi x_q \cos\theta_x/\lambda,
\end{equation}
where ${\phi}_q'$ is the QD phase received at the receiver $\hat{R}_x$.  Consequently, the received signal can be expressed as 
\begin{eqnarray}
\begin{array}{l}\label{12}
\hat{R}_{\theta_x} = h_{\theta_x} e^{j(\phi_y+\phi_q')}
	 + n_0 = h_{\theta_x} e^{j\phi_x+ j\Delta{\phi}_{x}} + n_0,
\end{array}
\end{eqnarray}
where $h_{\theta_x}$ and $\Delta{\phi_x} = 2\pi x_q (\cos\theta_x-1)/\lambda$ are the channel gain factor and the phase-offset owing to the geometric deviation angle $\theta_x$, respectively. 

Secondly, for the deviated perpendicular channel as shown in Fig.3(b), the received signal can be expressed as (see the Appendix)
\begin{eqnarray}
\begin{array}{l}\label{13}
\hat{R}_{\theta_y} = h_{\theta_y} e^{j(\phi_y+ kx_qsin\theta_y)} + n_0 = h_{\theta y}, e^{j(\phi_y+\Delta{\phi_y})}+n_0
\end{array}
\end{eqnarray}
where $h_{\theta_y}$ and $\Delta \phi_y= kx_q\sin\theta_y / \lambda$ are the channel gain factor and the phase-offset owing to $\theta_y$, respectively.  

Applying the Tailor expansion to \eqref{12} and \eqref{13}, we can find that the phase offsets are limited by the geometry deviation as $|\Delta \phi_x|\le  \frac{\pi\theta_x^2}{2} $ and $ |\Delta \phi_y|\le \pi \theta_y $.   By comparison, the perpendicular channel is more sensitive to the deviation angle than the parallel channel, because $|\Delta \phi_y|$ converges in a linear manner, while $|\Delta \phi_x|$ with square, when the deviation angles approach zero.  
    
As to the transmission performance, the QD method does not degrade much in principle, when the maximum values of $|\Delta \phi_x| $ and $|\Delta \phi_y|$ are much smaller than the phase difference between two adjacent points in the signal constellation used in the communication. To solidify this part of the work, simulations will be performed in the next subsection.

Finally, we note that the assumption of zero switching-time in the theoretical model can be removed, because the phase detection is done by comparing the RF oscillations between the arriving signal and the local reference, both of which experience the same switching-time.  Thus, there is no relative phase change as compared to the case with zero switching-time. 

\subsection{Simulations}
In this subsection, simulation results are reported for investigating the performances of one symbol through two directions in the ideally orthogonal- and deviated geometric model. 

For simplicity, we use BPSK modulation at the signal source and two omni-directional antennas located at positions $x_0 = 0 $ and $x_1 =\lambda/2$ to construct the QD transmitter.  In addition, considering the crossroad scenarios where line of sight must exist from either the cars on the road or from people along the roadside to the traffic light, we neglect the multipath signals, because our antennas can be placed near the position of the traffic light.  Actually, we select the AWGN channel model to the simulations for testing the bit-error-rate (BER) performances on various SNR-values, 

The random phases $\phi_x$ and $\phi_y$ are generated at the QD transmitter by using Matlab 2019, and the Monte-Carlo method is used to simulate the receiver with perfect channel catheterization, where the noise power is set to zero in the pilots.  Then, hard decision is used in the symbol demodulation.   

\begin{figure}[!t]
	\centering
	\subfigure[]{
		\centering
		\includegraphics[width=0.42\textwidth]{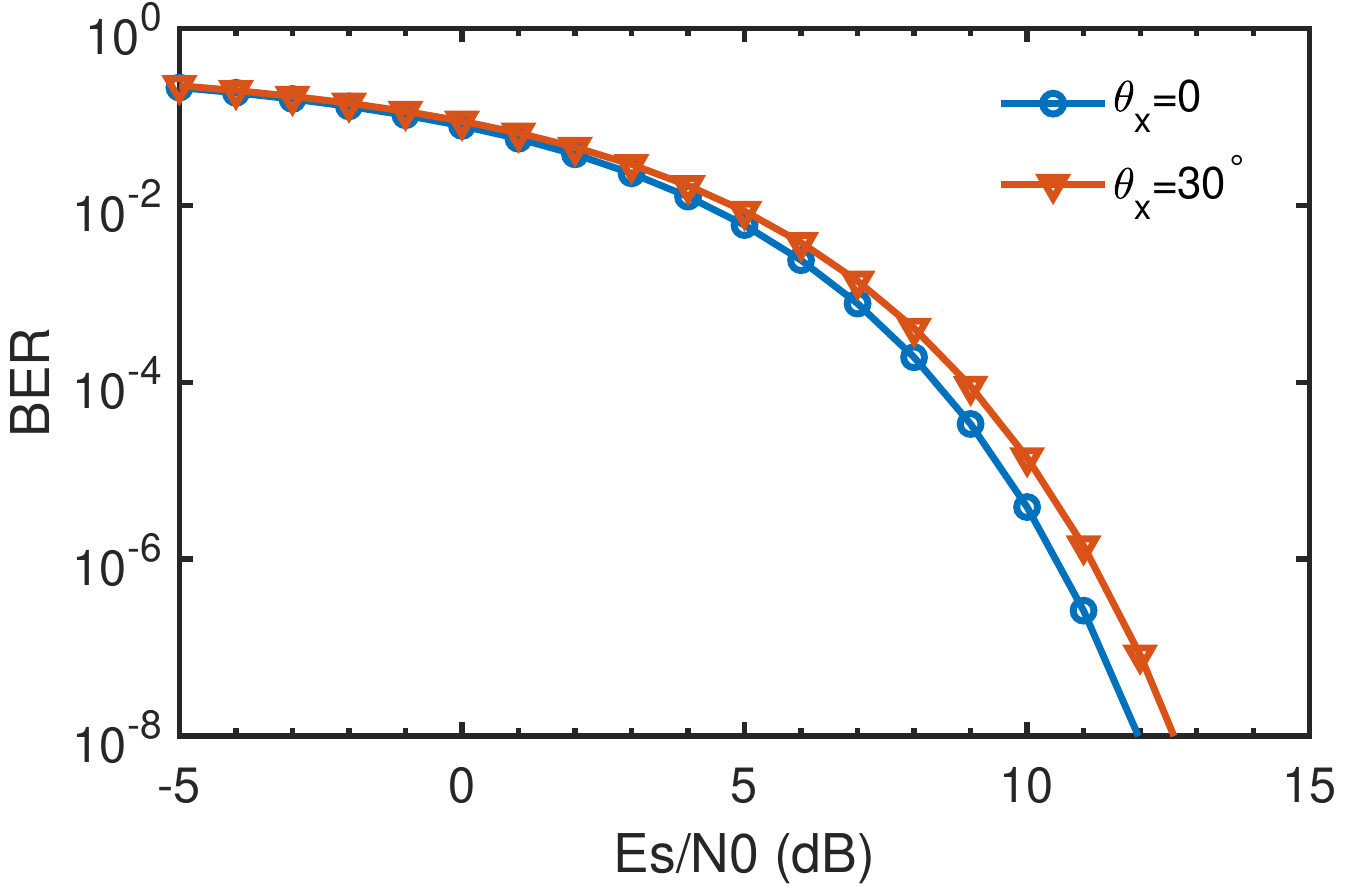}
		\label{fig4a}
	}
	\subfigure[]{
		\centering
		\includegraphics[width=0.42\textwidth]{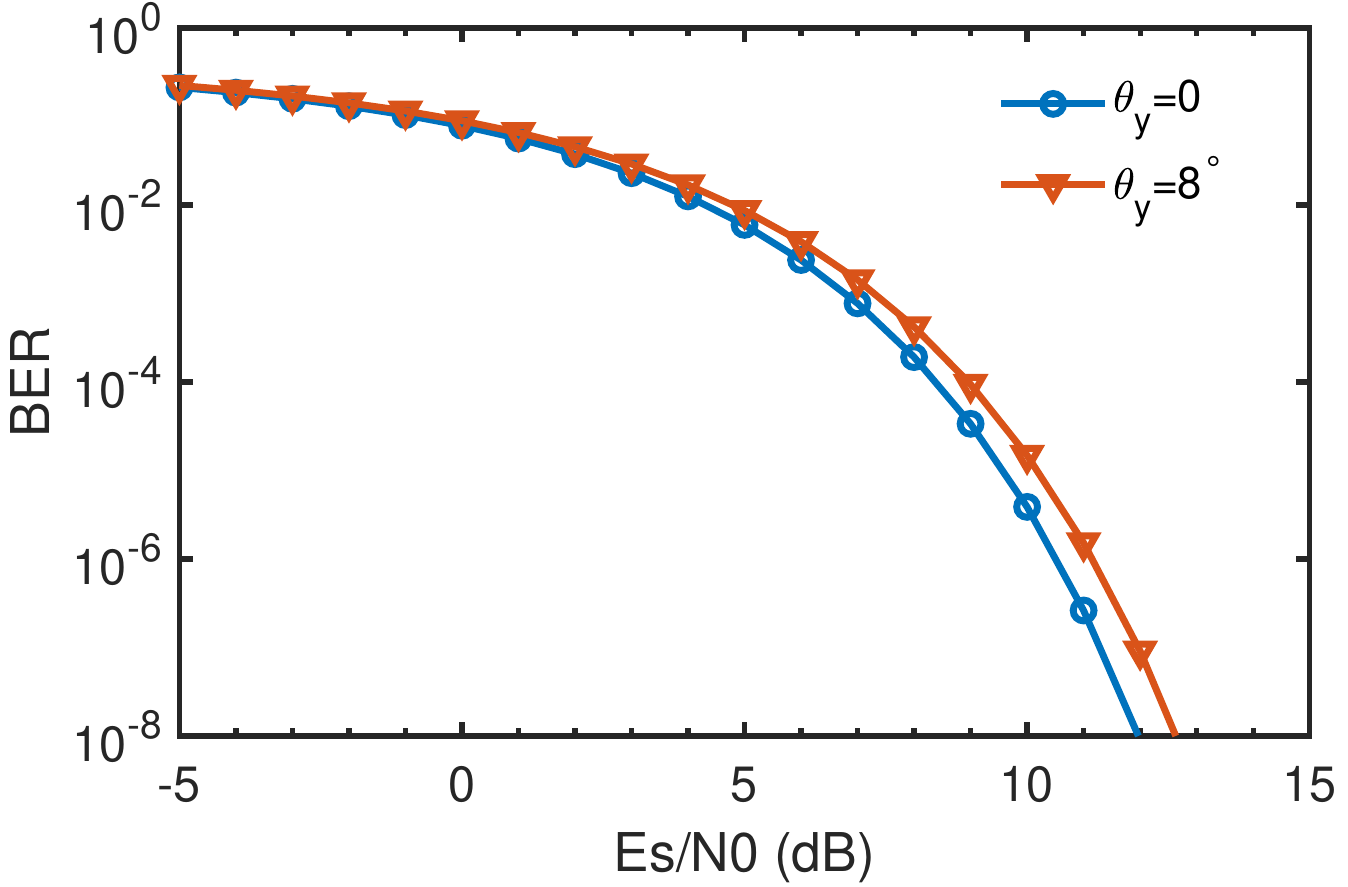}
		\label{fig4b}
	}
	\caption{The BER performance of the ideal channel and the deviated channel: (a) the parallel channel and (b) the perpendicular channel. }
	\label{BER_Performance}
\end{figure}

To confirm the QD method, the ideal geometric model is taken to the simulations and the results are plotted in Fig. 4(a)(b) for the parallel- and perpendicular channel, respectively.  Then, by setting a tolerable SNR loss at 0.8dB with $BER=10^{-8}$ in comparison with the ideal geometric model, we make a large number of simulations and found the corresponding deviation angles to $\theta_x = 30^o$ and $\theta_y = 8^o$, for which the BERs are also shown in Fig. \ref{BER_Performance}(a)(b), respectively. The relativity of the two deviation angles, i.e., $\theta_y < \theta_x$, agrees with the theoretical conclusion in the analysis, i.e. the perpendicular channel is more sensitive compared to the parallel channel.  

Finally, the author would get rid of any possible confusion between the QD method and the spatial modulation (SM) by the following short statement: the QD method can double the transmission efficiency and work in the one-to-two system, while the SM can increase the transmission efficiency slightly \cite{Yang2008} in the point-to-point manner \cite{Mesleh2008,Rajashekar2013,Wen2014}. 

\section{Conclusion}
The concept of the QD effect has been created for realizing the discrete phase modulation.  By combining the anisotropy of electromagnetic wave propagation, we developed the QD method that can double the transmission efficiency.  The application issues are addressed in the scenarios of crossroads with the ideal and deviated channel models, wherein the feasibility is confirmed by simulation results. 

\section*{Appendix}
By considering the plane wave model, we use Fig. A to show the angle deviations between the ideal- and deviated geometric model by $\theta_x$ and $\theta_y$. 

Examining the angle deviation from the parallel channel, one can find that the wave propagation path-difference $x_q -x_0$ in the ideal  geometric model should be projected onto the line in the direction of $\hat{R}_\theta$.  The projected result is equal to $(x_q-x_0) cos\theta_x$ that leads to the QD phase $kx_q cos\theta_x$ at the receiver. 
 
For the deviated perpendicular channel, the propagation path-difference is described by the deviation angle $\theta_y$.  Consequently, the QD phase can be changed to $kx_q sin\theta_y$ because $ cos\theta_x = sin\theta_y$.     
\renewcommand{\thefigure}{\Alph{figure}}
\setcounter{figure}{0}
\begin{figure}[!t]
	\centering
	\includegraphics[width=0.4\textwidth]{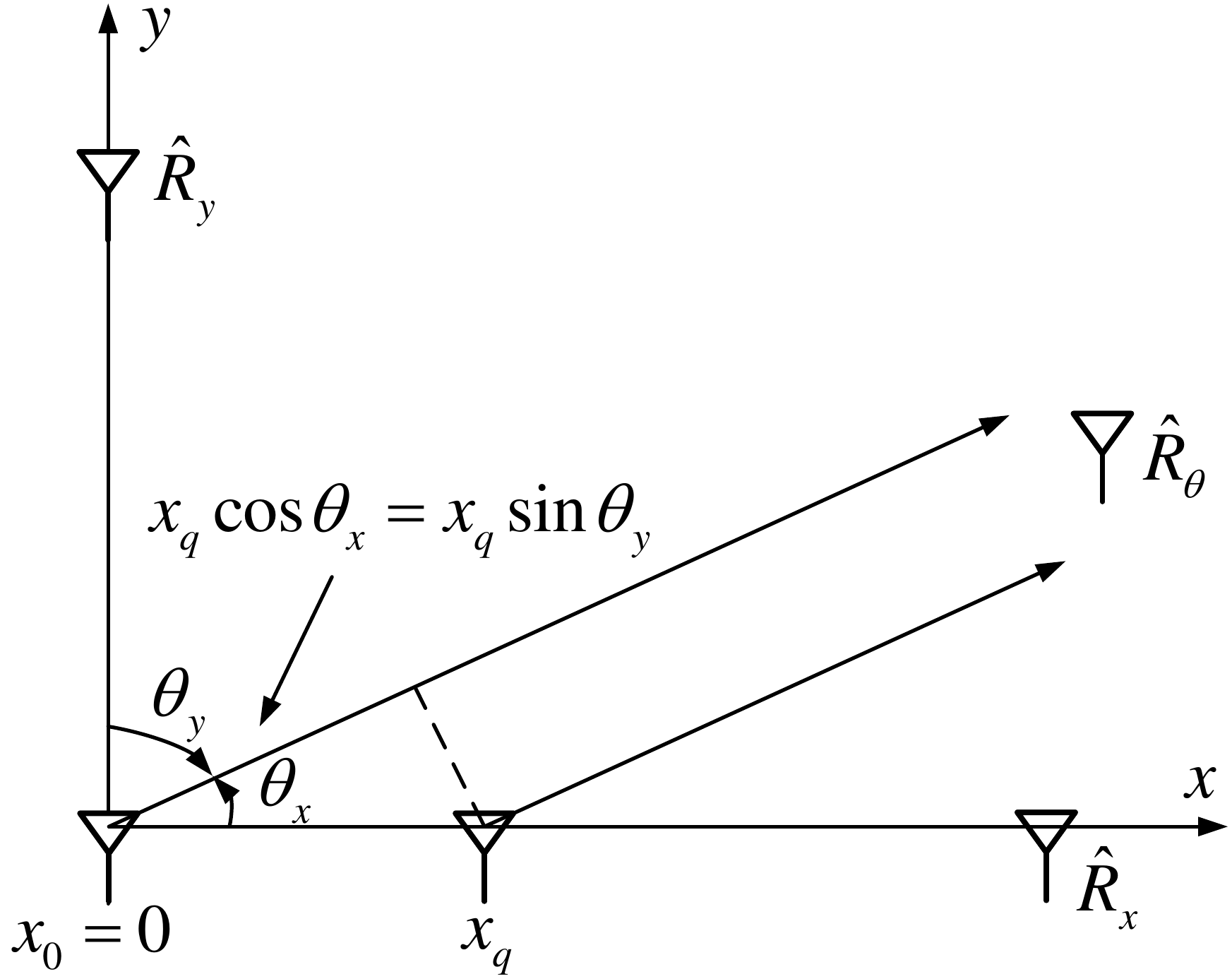}
	\caption{Illustration of the QD phase difference. }
	\label{appendix}
\end{figure}
\section*{Acknowledgment}

The author expresses his thanks to Dongsheng Zheng, a Ph.D candidate, for his work on the simulations. 

\bibliographystyle{IEEEtran}

\end{document}